\documentclass[%
 reprint,
 amsmath,amssymb,
 aps,
]{revtex4-2}
\DeclareUnicodeCharacter{2061}{}
\DeclareUnicodeCharacter{00A0}{}
\usepackage{graphicx}
\usepackage{dcolumn}
\usepackage{bm}

\begin{document}

\preprint{APS/123-QED}

\title{Floquet Engineering of Polaritonic Amplification in Dispersive Photonic Time Crystals}

\author{Mustafa Goksu Ozlu }
\thanks{These authors contributed equally to this work}
\author{Vahagn Mkhitaryan} 
\thanks{These authors contributed equally to this work}
\author{Colton B. Fruhling}%
\author{Alexandra Boltasseva}%
\author{Vladimir M. Shalaev}%
\affiliation{%
 Elmore Family School of Electrical and Computer Engineering \\ Birck Nanotechnology Center, Purdue University, West Lafayette, IN 47907, USA
}%




\date{\today}

\begin{abstract}
In this study, we investigate the dynamics of dispersive photonic time crystals (PTCs) and their potential applications for controlling light-matter interaction. By employing the Lorentz-Drude model, we analyze theoretically and via numerical simulation the effects of periodic modulation of dispersion parameters, revealing the emergence of hybrid bandgaps from interaction of polaritonic branches with unique characteristics. Our study demonstrates that dispersive PTCs offer novel excitation channels and amplification possibilities, that require lower modulation frequencies compared to non-dispersive systems thus alleviating experimental challenges for the realization of PTCs in the optical regime. These findings pave the way for advancements in polaritonic lasing and resonant Raman scattering. 
\end{abstract}

\maketitle


\section{\label{sec:level1}Introduction}
The study of time-varying photonic media has a history spanning several decades \cite{Louisell1961QuantumI.,Cassedy1963DispersionInteractions, Mollow1967QuantumI,Zurita-Sanchez2009Reflectiont}. Nevertheless, recent years have witnessed a resurgence of interest in this field revealing numerous new physical phenomena \cite{Galiffi2022PhotonicsMedia} such as the creation of squeezed and entangled photon states \cite{MendoncaQuantumRefraction, Dodonov1993QuantumMedia, Dodonov1998DynamicalStates, Mendonca2008VacuumAcceleration}, spatiotemporal beam steering \cite{Shaltout20153918Gratings, WangSpace-TimeWaves}, and the breaking of Lorentz reciprocity \cite{Sounas2017Non-reciprocalModulation, Chamanara2017OpticalGaps}. One of the most notable outcomes of engineering temporal dynamics is the concept of photonic time crystals (PTCs)\cite{Zurita-Sanchez2009Reflectiont,10.1063/1.4928659}, where the permittivity of the host material undergoes a periodic modulation with time, analogous to spatial photonic crystals which are periodically patterned in real space \cite{John1987StrongSuperlattices,Yablonovitch1987InhibitedElectronics}. Thanks to the non-Hermitian properties of PTCs these systems exhibit unique light-matter interactions, leading to intriguing phenomena such as generation and amplification of radiation fields from free electrons and dipoles \cite{Lyubarov2022AmplifiedCrystals,Dikopoltsev2022LightTime-crystals}.

Abrupt and strong changes in material's optical properties over time result in reflections and refractions at temporal interfaces \cite{MendoncaTemporalInterference,Shaltout2016PhotonicBand-Gaps}, akin to spatial boundaries. However, unlike spatial boundaries, temporal ones require dynamical modulation of optical properties, creating an open system and allowing time-reflected and refracted waves to have different energies than the incident beam. PTCs exploit this to create and amplify electromagnetic radiation in their momentum bandgap region, where time-reflected and refracted waves constructively interfere \cite{Pendry2021GainMechanism,Lyubarov2022AmplifiedCrystals}. This amplification resembles parametric amplification \cite{Mollow1967QuantumI,Louisell1961QuantumI.,10.1063/5.0091718,Cartella2018ParametricPhonons} but differs in two distinct features.  Firstly, the presence of strong modulation of material parameters, such as strongly modulating permittivity around the epsilon near zero point\cite{Kinsey2015Epsilon-near-zeroWavelengths, Reshef2019NonlinearMedia}, breaks the applicability limits of the conventional perturbative nonlinear treatment. Secondly, it is assumed that the entire medium undergoes simultaneous modulation while the probe beam propagates inside the medium, relaxing the typical phase-matching conditions \cite{KhurginPhotonicOn}.

Amplification within the bandgap of PTCs fundamentally arises from the transfer of energy from the modulation to the probe. This process necessitates interactions among various frequency components within the medium, leading to the creation of new components at the probe frequency. In a non-dispersive medium, where only a single frequency mode is supported for each momentum, effective amplification requires that the modulation frequency be twice that of the probe beam. Thus, accessing the momentum bandgap in non-dispersive media demands modulation at twice the probe beam frequency, which poses significant challenges for the realization of PTCs at optical or near-infrared (NIR) frequencies. This is why time reflection and exponential growth in the momentum bandgaps have so far only been observed experimentally in the microwave domain \cite{Wang2023Metasurface-basedCrystals,Moussa2023ObservationInterfaces}. However, recent advances in optical materials that can be strongly modulated on an femtosecond time scale \cite{Kinsey2015Epsilon-near-zeroWavelengths, Alam2016LargeRegion, Lustig:23,Saha:23} create a potential for experimental realization of PTCs also in the optical domain \cite{Lustig2023Time-refractionModulation,Lustig:21}. These materials are typically strongly dispersive and lossy. The material dispersion enforces electric field to evolve continuously in time, and potentially makes it more challenging to observe unique PTC behavior\cite{Hayran2022hbarCrystals}.  Additionally, the dispersive nature of materials couples energy from the light to mechanical motion in the material forming hybrid states of light and matter known as polaritons, which are new quasiparticles of the coupled system with well-defined energy and momentum. Building on the nondispersive foundations of recent PTC studies \cite{Lyubarov2022AmplifiedCrystals,Dikopoltsev2022LightTime-crystals,Felsen1970WaveMedia,Cassedy1963DispersionInteractions,Pendry2021GainMechanism}, we explore the inclusion of material dispersion in our analysis which opens up the possibility of investigating the interaction between matter and radiation field in the presence of optical modulation. We further discuss and examine more realistic experimental scenarios exploiting the full potential of such materials.

Recent studies for this exciting prospect have begun by investigating plane wave propagation in a dispersive time-varying media \cite{Solis2021Time-varyingDiscontinuity}, a two-level model with periodic modulation of resonance frequency \cite{Sloan2020DispersionCrystals,Lopez2022AtomicElectrons} and widening of momentum bandgap near material or structural resonance \cite{Wang2023UnleashingSystems}. A broader investigation of polaritonic interactions and its effect on PTC amplification is still needed. In this paper we focus on analyzing the overall PTC dispersion, also referred to as optical band-diagram, in the presence of Lorentz-Drude type material dispersion. Specifically, we explore the modulation of the coupling between different polaritonic branches. We observe the emergence of hybrid bandgaps which stems from interaction between upper and lower polaritonic branches. These hybrid bandgaps have experimentally favorable conditions. 

Our main findings reveal that when the parameters of dispersive materials are modulated, additional bandgaps with lower modulation speeds emerge in the system. These bandgaps are accessible from the modes of the unmodulated medium and enable excitation channels for coherently interfering time-reflected and refracted waves, establishing exponentially growing fields. This relaxes the fast modulation requirement of nondispersive materials and suggests more accessible experimental pathways for observing such effects with much slower modulation frequencies. 

As amplification in PTCs stems from non-perturbative nonlinear interactions, the amplification of waves within the polaritonic PTCs is intricately connected to the nonlinear interaction among polaritons \cite{Mills1974Polaritons:Media}. Thus, by examining polaritonic PTCs we may gain valuable insights into the physics of these strong, non-perturbative interactions among polaritons that pave the way forward for enhanced resonant Raman scattering and polaritonic lasing \cite{Kena-Cohen2010Room-temperatureMicrocavity,Gorelik1969RAMANCRYSTALS}.

\section{\label{sec:level1}Polaritonic Bands with Modulation}

When the optical properties of a medium are periodically modulated with a frequency $\Omega$, its properties remain time-translation invariant under a temporal shift of $T = 1/\Omega$, i.e. $\Psi(t+T)=\Psi(t)$, where $\Psi$ is a physical quantity such as electromagnetic field amplitude or polarization in the medium and t is the time. This temporal periodicity causes the $\omega-k$ dispersion of the supported modes to also become periodic in frequency with period $\Omega$. Therefore, the dispersion of the system is completely characterized by the frequency range of 0 to $\Omega$, known as the 1st Brillouin zone.  Owing to this periodicity, the modes of the medium can be described using the Floquet theory, by expanding the relevant quantities in the form $\boldsymbol{\Psi}(t)=e^{i\omega_{k}t}\sum_{n=-\infty}^{\infty}\boldsymbol{\Psi}_{n}e^{in\Omega t}$
,with $\omega_{k}+2\pi n=\omega_{k}$, where ${n}$ is an integer. 

\begin{figure}[h]
\includegraphics[width=0.5\textwidth]{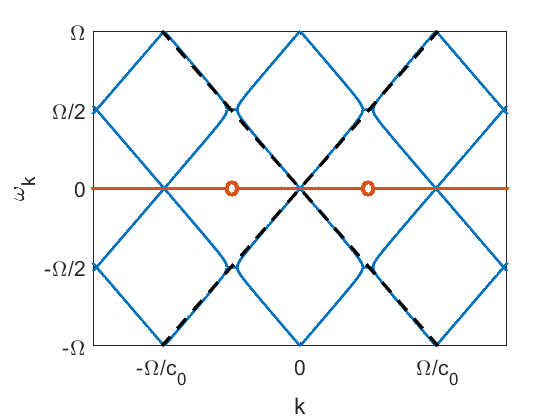}
\caption{\label{fig:nondispersive}(a) The band diagram of a non-dispersive photonic time crystal with electric permittivity $\varepsilon=1+0.3\sin{⁡(\Omega t)}$ the dashed green line shows the dispersion relation of the unmodulated medium with $\varepsilon=1$. The blue lines indicate the real parts of eigenvalues  while the red indicates the imaginary parts.}
\end{figure}

In the case of a stationary, nondispersive optical medium, the dispersion relation is given by the expression $\omega=kc/n$, where c is the speed of light in a vacuum, and n is the refractive index. This relation defines a single line for positive $\omega$ and k shown in Figure \ref{fig:nondispersive} by black dashed lines. The effect of periodic modulation of the dielectric function on the dispersion relation for such nondispersive materials can be visualized as dispersion lines of the unmodulated medium being translated along the frequency axis by integer multiples of the modulation frequency $\Omega$ . In this case the dispersion for negative frequencies also becomes relevant as they get shifted to the positive frequency domain and mix with other dispersion lines. The translation of the dispersion lines causes the branches in different Brillouin zones to cross each other leading to complex frequency eigenvalues, $\omega_k$, with degenerate real parts. This contrasts with the unmodulated, lossless medium, which has only real eigenfrequencies. Following the analogy of the frequency band gaps in spatial photonic crystals, these regions with imaginary eigenfrequencies have been called momentum bandgaps with the characteristic signature of exponentially growing photon fields \cite{Lyubarov2022AmplifiedCrystals,Dikopoltsev2022LightTime-crystals}. The emergence of the momentum bandgaps and complex eigenvalues are illustrated for a nondispersive medium in Figure \ref{fig:nondispersive}. The modes, as they are linear, cross each other at frequency $\Omega/2$ resulting in a bandgap at that frequency. 
\begin{figure*}
\includegraphics[width=\textwidth]{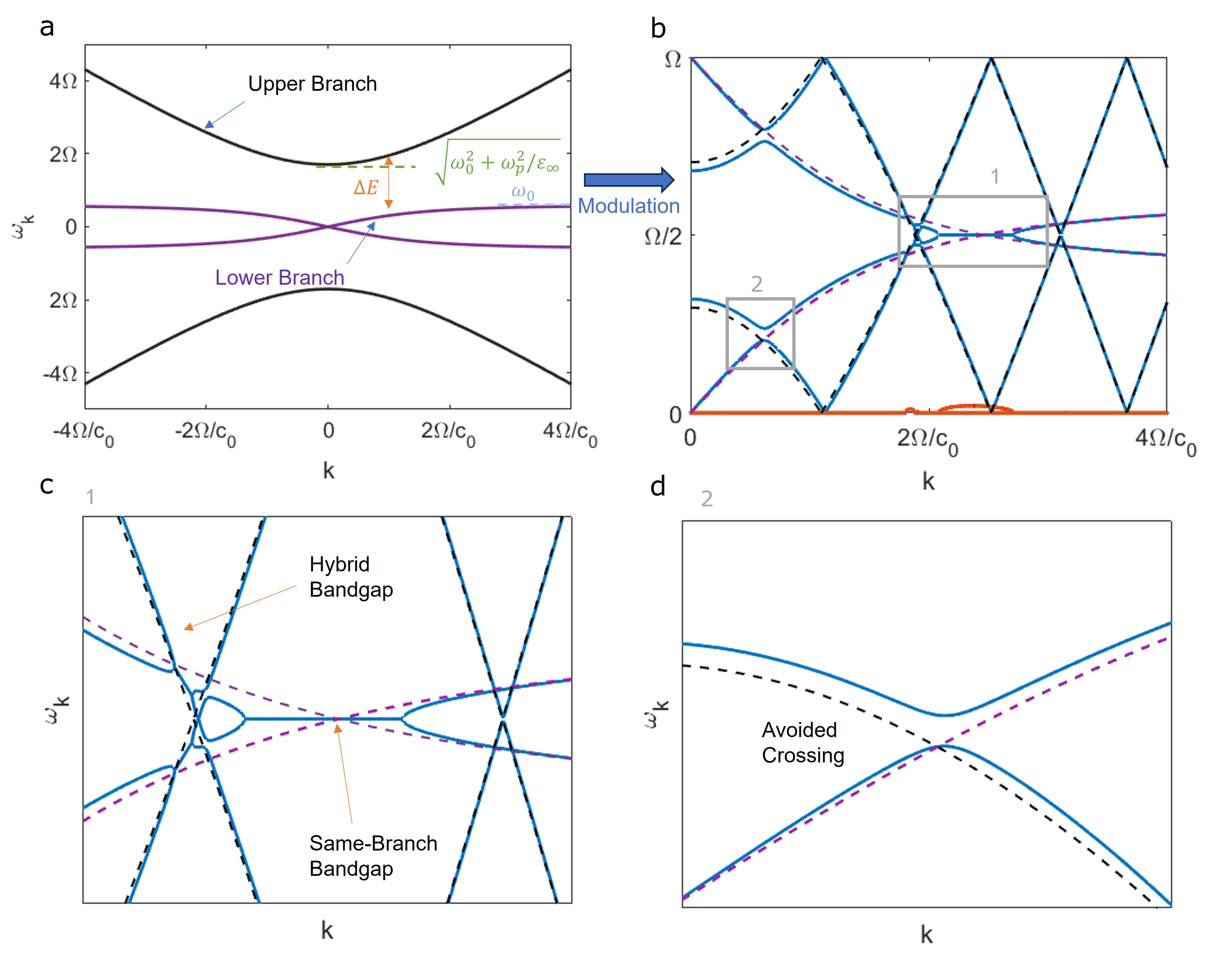}
\caption{\label{fig:effect}(a) The band diagram of a non-dispersive photonic time crystal with electric permittivity $\varepsilon=1+0.3\sin{⁡(\Omega t)}$ the dashed green line shows the dispersion relation of the unmodulated medium (b) The dispersion diagram of a Lorentz type material with $\omega_p=1.6\Omega$  $\omega_0=0.6\Omega$ $\varepsilon_\infty=1$ (c) PTC band diagram of the dispersive material presented in part b. The plasma frequency is modulated with $\omega_p (t)=1.6+0.8\sin{⁡(\Omega t)}$. The blue lines show the real and the red lines show the imaginary part of the frequency eigenvalues. The dispersion branches of the unmodulated medium is presented with dashed lines (d) A close up of the region where two types of momentum bandgaps are highlighted (e) A close up of the region where the avoided crossing is highlighted.}
\end{figure*}

In the case of a dispersive medium, the energy no longer remains purely in the fields. The electromagnetic waves hybridize with material excitations and form polaritons. The response of the materials that possess bound charges can be described classically with Lorentz type frequency-dependent dielectric function.
\begin{eqnarray}
\varepsilon(\omega)=\epsilon_{\infty}+\dfrac{\omega_p^2}{\omega_{0}^{2}-\omega^{2}-i\gamma\omega}
\label{eq:one}
\end{eqnarray}
where, $\omega_0$, $\gamma$, and $\varepsilon_\infty$ are respectively the central frequency, Drude damping and high-frequency permittivity of the material and $\omega_p^2=4\pi e^2N/m^*$ is the plasma frequency of the material defined by free carrier concentration N and effective mass $m^*$. This form of dielectric function follows from the harmonically oscillating polarization density of the material excited by electromagnetic waves of the same frequency.
\begin{eqnarray}
\frac{d^{2}P}{dt^{2}}+\gamma\frac{dP}{dt}+\omega_{0}^{2}P=\frac{Ne^{2}}{m^*}E
\label{eq:two}
\end{eqnarray}
Polaritons in such media are formed when electromagnetic waves couple to the mechanical oscillations of these bound charges through induced polarization created due to their harmonic motion. The dispersion relation for such material follows from Maxwell’s equations and is given by ${k^2=\omega^2\varepsilon(\omega)/c^2}$. The solution of this quadratic equation leads to two dispersion branches above and below characteristic frequency $\omega_0/\sqrt{\varepsilon_\infty}$, which are conventionally called upper and lower polaritonic branches. These two branches arise from the coupling of phononic and photonic dispersion lines. Due to the coupling, the branches bend outward and form avoided crossing with an energy difference $\Delta E$ known as the Rabi splitting. An example of such dispersion curves is shown in Figure \ref{fig:effect}.a calculated for normalized oscillator parameters $\omega_p=1.6\Omega$, $\omega_0=0.6\Omega$, and $\varepsilon_\infty=1$ .

We now incorporate temporal changes in material properties using dispersion parameters. Although our model includes dispersive effects the response of dispersion parameters to external stimuli is still assumed to be instantaneous. Experimentally the most accessible modulation is through optical pumping of semiconductors and changing conduction band electron density or effective mass. These two parameters directly affect the plasma frequency, therefore in our model, we assume to be modulating the plasma frequency.
\begin{eqnarray}
  \omega_p^2(t)=\omega_{p0}^2[1+\Delta^2\cos{(\Omega t)}]
  \label{eq:three}
\end{eqnarray}

Due to this periodicity, the fields inside the medium can be written as Floquet expansion and the following eigenvalue equation is obtained from Maxwell's equations.
\begin{eqnarray}
\omega_{k}\boldsymbol{\Psi}_{p}=\sum_{n=-\infty}^{\infty}(\boldsymbol{M}_{p-n}-n\Omega I_{4\times4}\delta_{pn})\boldsymbol{\Psi}_{n}
\label{eq:four}
\end{eqnarray}
where ${\bf\Psi}_p^k=[E_{xp},H_{yp},J_{xp},P_{xp}]^T$, with superscript T meaning transpose operation and ${J_{xp}=\partial_t P_{xp}}$, and matrices $\bf \bar{M}_m $ are given by

\begin{eqnarray}
  \boldsymbol{M}_{m}=\left
  [\begin{array}{cccc}
0 & -ck\delta_{m0}/\epsilon_{\infty} & 4\pi i\delta_{m0} & 0\\
-ck\delta_{m0} & 0 & 0 & 0\\
-if_{m} & 0 & i\gamma_{m} & i\omega_{om}^{2}\\
0 & 0 & -i\delta_{m0} & 0
\label{eq:five}
\end{array}\right]
\end{eqnarray}
The summation in Eq.\ref{eq:four} runs over all possible Floquet modes, however in numerical implementation, we consider only finite number modes and truncate this infinite system of equations into generalized eigenvalue problem:
\begin{eqnarray}
  \omega_{k}\boldsymbol{\Psi}^{(k)}=\boldsymbol{\bar{A} \Psi}^{(k)}
  \label{eq:six}
\end{eqnarray}
with ${\bf \Psi} = [{\bf\Psi_1^{(k)}},{\bf\Psi_2^{(k)}},...,{\bf\Psi^{(k)}}]$ and $\bf\bar {A}$ is a matrix formed by elements from Eq.\ref{eq:five}. 
Here for an isotropic medium, without losing the generality we assumed the waves propagate along the z-axis. The equations for waves with orthogonal polarization ${\bf\Psi}=[E_y,H_x,J_y,P_y]^T$ have very similar form, hence we will focus only on one of them for our discussions. The details of the calculation can be found in the Supplementary Information(SI).

Solving the eigenvalue Eq.\ref{eq:six} results in the band diagram for a dispersive photonic time crystal along with the corresponding eigenfunctions. Introducing modulation to a dispersive medium, like in the nondispersive case, causes the dispersion branches to be translated with the modulation frequency. The overlapping polaritonic branches interact and give rise to bandgaps or avoided crossing based on the type of dispersion line as shown in Figure \ref{fig:effect}.b. The dashed lines show the translated dispersion lines of the stationary medium and the blue lines show the dispersion line of the modulated medium. The dispersion parameters and eigenfrequency of the Floquet modes are expressed relative to the modulation frequency $\Omega$.

The presence of polaritonic bands and their modulation introduce novel phenomena into the physics of time-varying medium. One of the most prominent effects for this case is the interaction between lower and upper polariton branches. The crossing of lower and upper branches belonging to positive and negative frequency quadrants causes the real part of the frequency eigenvalues to split and give rise to a frequency bandgap highlighted in Fig \ref{fig:effect}.c which is otherwise not observed in nondispersive case. 

The interaction of polaritonic branches causes the emergence of bandgaps as well. First, a bandgap with similar characteristics to non-dispersive PTC bandgaps arises from the interaction between two lower branches, highlighted in Fig \ref{fig:effect}.d as same branch bandgap. These bandgaps occur in the middle of the unit cell at frequency $\Omega/2$, with a flat dispersion line. A distinguishing factor for these bandgaps from the non-dispersive case is the additional excitation channels made available by the upper branch with matching momentum. 

Furthermore, the crossing of upper and lower polaritonic branches leads to the formation of a novel bandgap which we term as hybrid bandgaps with distinct features. First, hybrid bandgaps does not occur on a flat line resulting in nonzero group velocity. Therefore, the energy is amplified directionally. Secondly, the modulation frequency required to create this bandgap is not necessarily twice the frequency of the unmodulated branch, but it is the energy difference between the upper and lower branches. Therefore, the requirement of having the modulation speed twice the probe beam's frequency does not apply. 

\begin{figure}[h]
\includegraphics[width=0.5\textwidth]{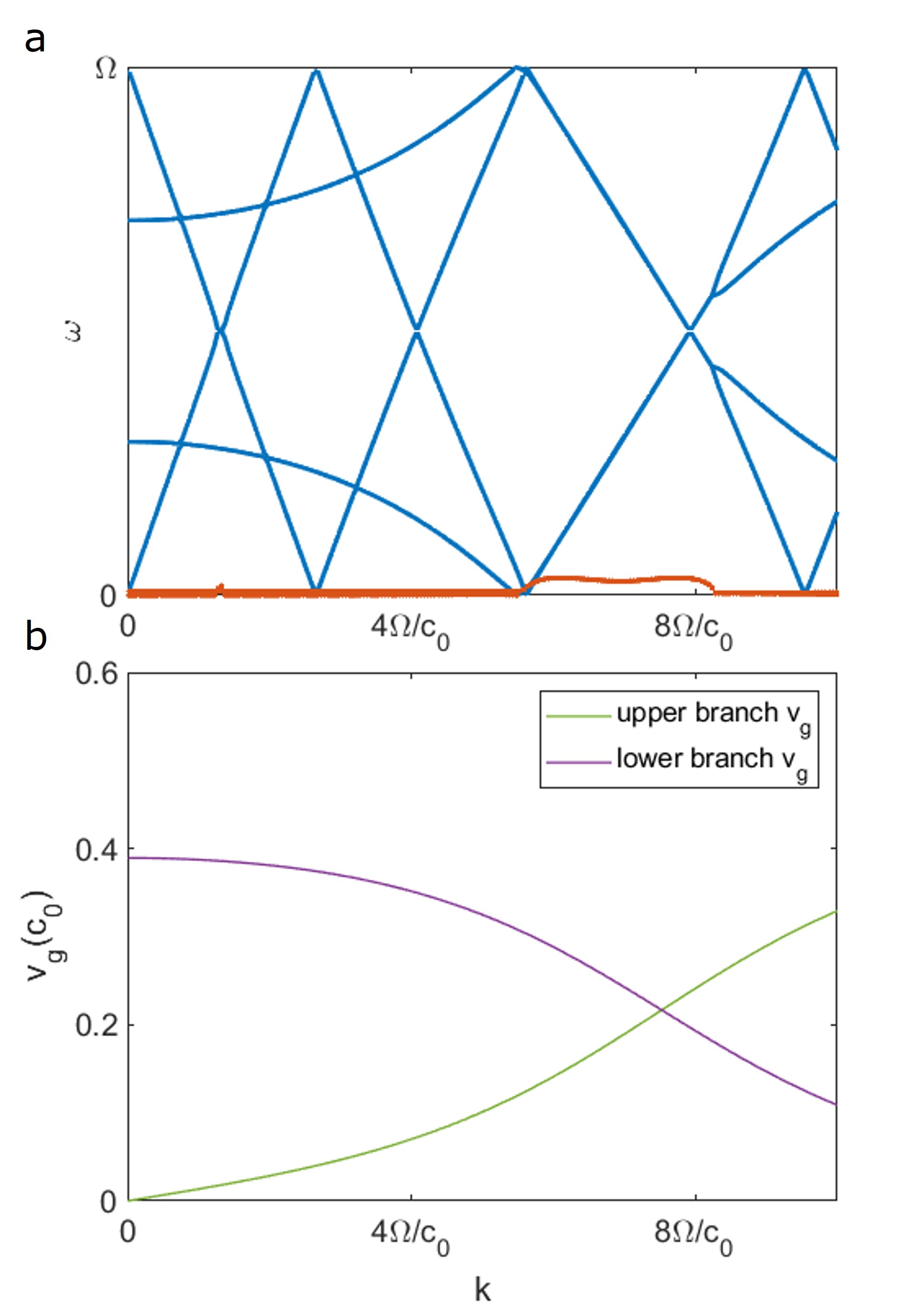}
\caption{\label{fig:hBN_mod} (a) PTC band diagram of the hBN like dispersive material presented in Table 1. The plasma frequency is modulated with $\omega_p(t) = 0.61+0.3\sin⁡{(\Omega t)}$ where $\Omega = 13 \rm THz$. The blue lines show the real and the red lines show the imaginary part of the frequency eigenvalues (b) Group velocity of the upper and lower branches for the unmodulated hBN dispersion}
\end{figure}

The hybrid bandgap's size and amplification heavily rely on the coupling strength between the upper and lower branches, and is maximized when the group velocities of the branches match to allow a more profound overlap. This alignment occurs when the hybrid bandgap coincides with the momentum region where Rabi splitting is present between the unmodulated branches. We accentuate this effect using dispersion parameters of a well-known phononic optical material hexagonal boron nitride (h-BN)(see Table \ref{tab:hBN})\cite{Caldwell2014Sub-diffractionalNitride}. The band diagram for this case is given in Fig \ref{fig:hBN_mod}.a.

\begin{table}[b]
\caption{\label{tab:hBN}%
 The dispersion parameters of hBN in tangential direction. Modulation frequency $\Omega$ = 12 THz
}
\begin{ruledtabular}
\begin{tabular}{c c c c}
\textrm{$\omega_p$}&
\textrm{$\omega_0$}&
\textrm{$\gamma$}&
\textrm{$\varepsilon_\infty$}\\
\colrule
58 THz & 41 THz & 0.2 THz & 4.9\\
1920 $\rm cm^{-1}$ & 1360 $\rm cm^{-1}$ & 7 $\rm cm^{-1}$ & \\
$2\sqrt{\varepsilon_\infty}$ $\Omega$ & 3.40 $\Omega$ & 0.017 $\Omega$ & 4.9\\
\end{tabular}
\end{ruledtabular}
\end{table}

For a phononic medium, the minimum separation of the lower and upper branches is equal to $\omega_p/\sqrt{\varepsilon_\infty}$ which is proportional to the difference between the transverse phonon (TO) and longitudinal phonon (LO) frequencies such that $\Delta \omega=\sqrt{\omega_{LO}^2-\omega_{TO}^2}$ [36]. The modulation frequency has been chosen to be half of the Rabi splitting ($\Delta \omega$) between the upper and lower branches. This allows maximal overlap coupling between lower and upper branches where their group velocities match, as shown in Fig \ref{fig:hBN_mod}.b, creating profound bandgaps. We investigate the characteristics of wave propagation in these bandgaps and the possibility of obtaining amplification with slower modulations in the subsequent section.

\section{Excitation of Modes}

For a given medium, the eigen solutions and their dispersions discussed in the previous section characterize the natural modes supported in that medium. When this medium is excited by external sources, the induced field can be represented analytically as time-dependent superpositions of these eigenmodes. Such analytical formulations give great flexibility to study the dynamics of the modes in the system. In particular, it allows us to observe how individual modes are excited, and how they evolve and interact among themselves. To demonstrate this, we discuss the dynamics of a dispersive, time-varying medium under dipole and plane wave excitations. 

\begin{figure}
\includegraphics[width=0.5\textwidth]{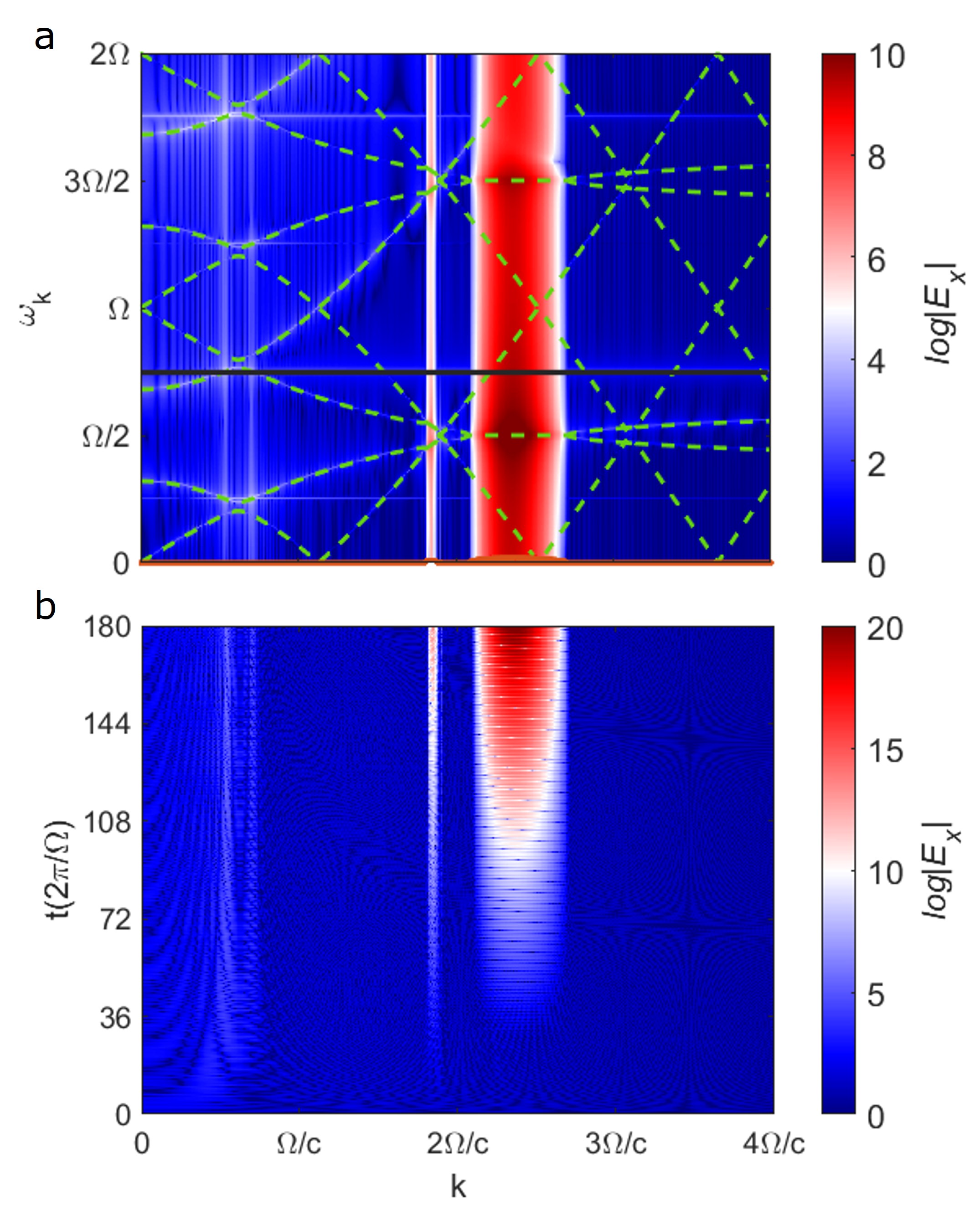}
\caption{\label{fig:dipole}(a) The excitation spectrum of the dispersive PTC discussed in Figure \ref{fig:effect}.c excited with a dipole source at frequency $0.75\Omega$ in log scale. The blue dashed lines show the real part of the band-diagram (b) Time-momentum graph showing the evolution of the fields in the excited medium in log scale.}  
\end{figure}

We first illustrate the system dynamics using a dipole excitation. This study builds on previous discussions regarding a non-dispersive medium \cite{Lyubarov2022AmplifiedCrystals}. Specifically, we examine the excitation of the system whose band diagram is depicted in Figure \ref{fig:effect}.b. The system is, excited by a time-harmonic dipole source with an oscillation frequency $\omega_d = 0.7\Omega$, chosen to fall outside of any bandgaps. Fig \ref{fig:dipole}.a shows the excitation spectrum as density plots superimposed by the band diagram (green dashed lines). 

The excitation frequency is represented by the black line. Initially, the light couples most strongly to the branches of the dispersion diagram closest to the excitation frequency. However, the entire band is also excited because the radiated light from a point source has all the momentum components. Additionally, higher-order bands are excited through modulation-induced interaction. Similar to the non-dispersive case, waves excited at the momentum bandgap grow exponentially and appear with the strongest amplitude.

Exponential growth can also be observed from the time-momentum graph in Figure \ref{fig:dipole}2b. Initially, the momentum values corresponding to the supported modes of the system at the source frequency are excited. As time progresses, the modes with the correct momentum range to couple into the bandgaps grow exponentially, eventually surpassing fields at any other wavenumber.

Excitation via dipole sources displays the range of modes. However, experimental excitation is usually confined to propagating or evanescent waves inside the medium, which are bound within the supported momentum range. To investigate excitation with a wave we employ FDTD simulations with a probe beam propagating within a modulated medium. The details of the calculation are provided in the SI.

Before modulation, the probe beam can have momentum and frequency residing only on one of the dispersion branches of the non-modulated medium. Once the modulation starts, the beam’s momentum remains constant while its frequency couples to the Floquet modes. When the probe beam momentum is within the bandgap, the beam couples into the momentum gap mode and starts growing. For a nondispersive medium,
the bandgap always occurs at $k=\Omega/(2n_{\rm avg}c) $ \cite{Dikopoltsev2022LightTime-crystals,Lyubarov2022AmplifiedCrystals}. In such a medium, for a wave to have this momentum, its frequency should always be $\Omega/2$ Hence, exciting the bandgap requires very fast and strong modulation of the optical properties, making it challenging to possibly observe photonic time crystals in the optical regime. However, the situation is dramatically different for dispersive materials and these strict conditions can be relaxed as there are multiple features, including new types of band gaps and avoided crossings in the band diagram that lead to different excitation regimes.

\begin{figure*}
\includegraphics[width=\textwidth]{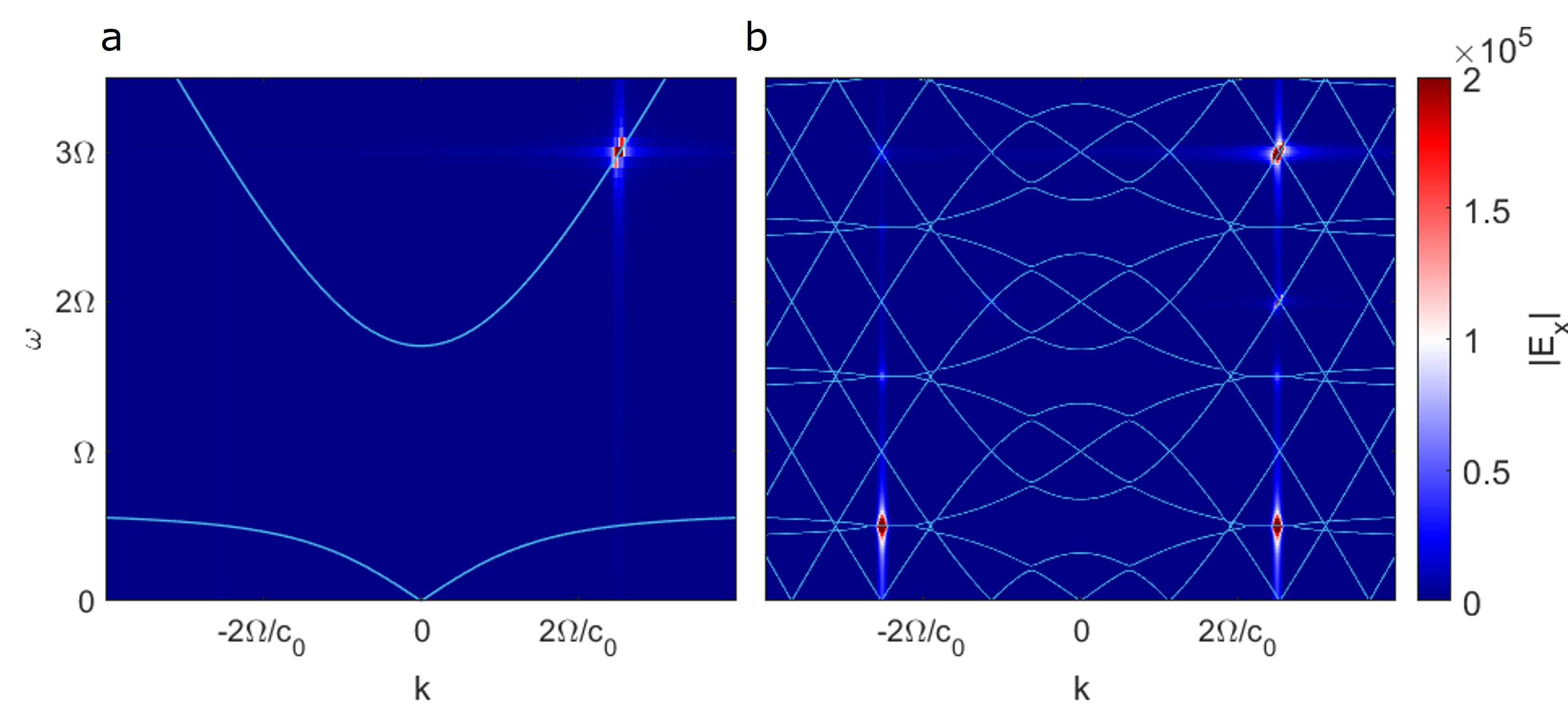}
\caption{\label{fig:same}(a) Frequency-momentum relation of the excitation wave inside the dispersive medium without modulation. The blue lines show the dispersion relation of the unmodulated medium. (b) The frequency-momentum relation of the field inside the dispersive PTC. The blue lines show the band-diagram of the PTC}  
\end{figure*}

 First, we investigate the same-branch bandgaps. We again use the dispersion parameters of the medium in Figure \ref{fig:effect}.c. In this case the bandgap occurs at 2.5 $k=\Omega/c$. Additionally, contrary to the non-dispersive case there are two frequencies where propagation with this momentum is allowed for the non-modulated medium, on the upper and lower branches. 
We excite this medium from the upper branch with $\omega= 3\Omega$ as highlighted in Figure \ref{fig:same}.a. 
Once the modulation starts some of the energy remains at the original branch and continues to propagate without a change in the amplitude. This is a result of the original branch being almost identically present in the dispersion of the modulated medium and the bandgap does not occur on this branch. 
As shown in figure \ref{fig:same}.b some portion of the energy of the incident beam couples into the bandgap created by the lower branches occurring at $\Omega/2$ and at every integer multiple of $\Omega/2$. The light in this branch shows the same characteristics as the bandgap mode in a nondispersive medium. As the modulation starts, the beam coupled to the bandgap is equally separated into forward and backward propagating modes and grows exponentially. The bandgap occurs on a flat band, therefore, the group velocity of the beam is zero and the wave remains localized in space until modulation ends. It is observed that the new modes are excited in both quadrants equally, which means the forward and backward propagating beams have equal amplitudes. The video of the temporal evolution of the beam for this excitation is given in Supplementary Video 1. Higher order frequencies are also visible at $3\Omega/2$ as a result of harmonic generation with the modulation. Although this case demonstrates transferring energy between polaritons of the medium, the bandgap mode still occurs at half the frequency of the modulation.

\begin{figure*}
\includegraphics[width=\textwidth]{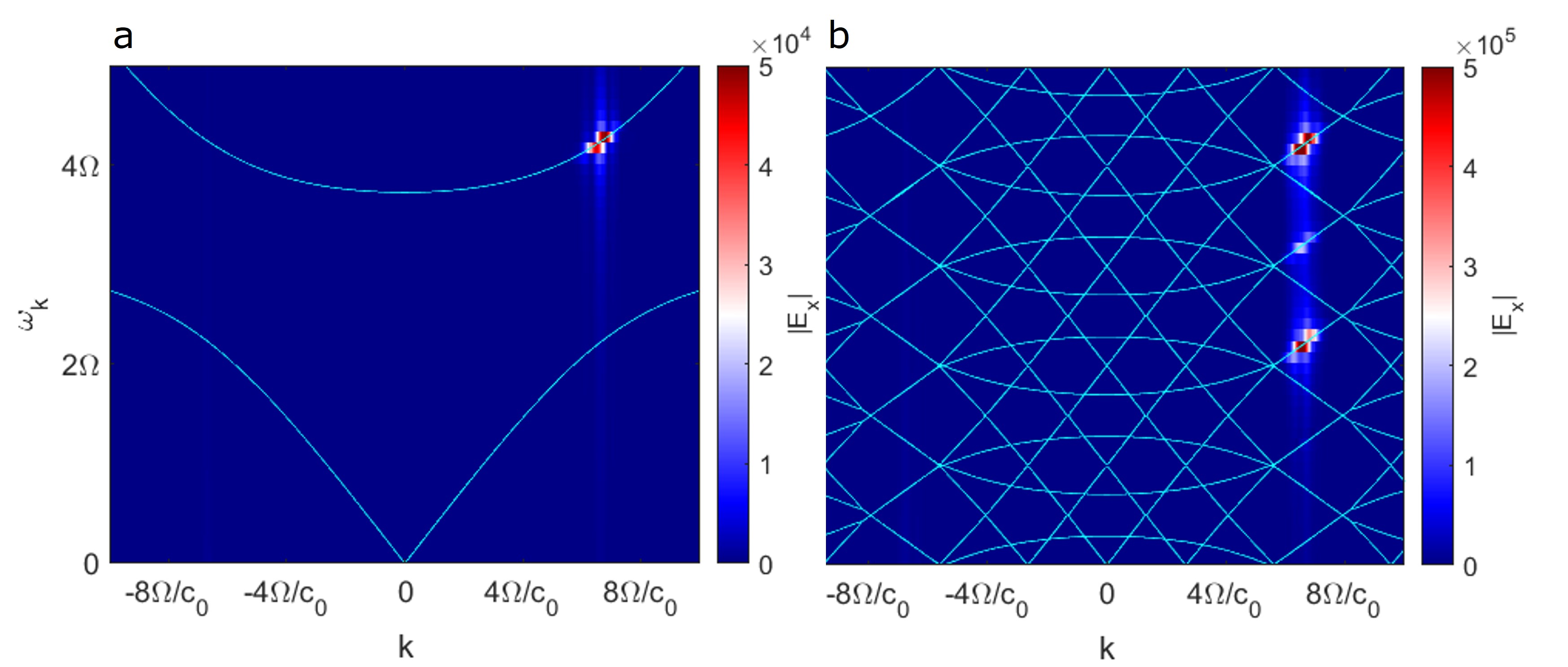}
\caption{\label{fig:hybrid}(a) Frequency-momentum relation of the excitation wave inside the medium with hBN like dispersion without modulation. The blue lines show the dispersion relation of the unmodulated medium. (b) The frequency-momentum relation of the field inside the dispersive PTC. The blue lines show the band-diagram of the PTC.}  
\end{figure*}

This is not the scenario for the hybrid bandgap. We investigate the band diagram shown in Figure \ref{fig:hBN_mod}.a. The hybrid bandgap occurs around $ k=7\Omega/c$ where the upper and lower branches cross each other. To couple into the bandgap mode, we excite from the upper branch at $\omega = 4.23\Omega$ and $k =6.7\Omega/c$ as shown in Figure \ref{fig:hybrid}.a. After modulation is introduced, the bandgap occurs on the branch where the beam was already propagating before the modulation, therefore all the energy directly couples to the bandgap mode resulting in amplification. Notably, the bandgap in this case is not a flat band as in the same-branch case but is instead slanted. Therefore, the resulting amplified beam propagates with non-zero group velocity. This is a distinct difference between the dispersionless cases investigated previously. Figure \ref{fig:hybrid}.b shows the $\omega-k$ relationship of the beam inside the medium during the modulation where one can observe the growth in the field amplitude and the excitation of the lower branches $\Omega$ away in frequency from the probe beam. Supplementary Video 2 shows the temporal evolution of this excitation.

One notable observation of this structure is the requirement of modulation speed is more than 8 times lower than that of a non-dispersive medium. The interaction between the polaritonic branches allows for PTC bandgap-like amplification to happen at significantly slower modulation. In the example shown a modulation at 13 THz gives rise to directional amplification of a beam at 55 THz. This feature is a clear path forward for observing PTC bandgaps in the infrared (IR) regime and possibly in the optics regime with appropriate selection of material and modulation properties.

\section{Conclusion}

In this work, we have investigated the effects of periodic temporal modulation on dispersive mediums. The dispersion lines of the material start translating and mixing via modulation. The resulting band diagrams exhibit bandgaps that emanate from the interaction of polaritonic branches. Due to dispersive materials being able to support several modes at a given momentum, there are more possibilities for interactions. This allows for the occurrence of new bandgaps and new channels for excitation. 

One significant feature observed in the polaritonic PTCs is the hybrid bandgap that occurs through the interaction of a lower and an upper dispersion branch. The region where the two branches have the closest energy and exhibit avoided crossing is where they are most strongly coupled. When driven by their energy difference in this regime, it is possible to create a strong interaction between polaritons and the modulation, that allows for energy to be transferred from modulation to the probe beam. This is achieved with much lower modulation frequencies than what is necessary for non-dispersive PTCs. Allowing for experimental observations in higher energy regimes of the electromagnetic spectrum.

When modulated with a frequency proportional to the energy separation of the LO and TO phonons, phononic materials can be used for polaritonic amplification. Although creating strong optically induced changes in these materials are challenging, it has been shown that it is possible to drive nonlinear response in phononic materials through Born effective charge and dielectric screening of the electric field by electrons. This nonlinearity has been used to observe parametric amplification of optical phonons.\cite{Cartella2018ParametricPhonons} 

Apart from employing the phononic material dispersion one other possible realization of polaritonic amplification is through constructing the desired dispersion lines through structured materials. Bulk and surface plasmons of thin films hybridize near the epsilon near zero region of the thin films to give rise to Berreman and ENZ modes, which can be coupled through relatively low frequency modulation to give rise to amplified polaritons.
The interaction of polaritonic modes in dispersive photonic time crystals (PTCs) has the potential for amplification across a broad electromagnetic spectrum. This enhancement can impact the scope and applications of polaritonic lasing, resonant Raman scattering, and polariton nonlinearities.
\begin{acknowledgments}
Authors acknowledge support by the U.S. Department of Energy under Award DE-SC0017717 (Control of Light), and the Air Force Office of Scientific
Research under Award FA9550-21-1-0299
\end{acknowledgments}

\bibliography{mendeley}
\end{document}